\documentclass{article}
\usepackage{graphicx}

\def\be{\begin{equation}}
\def\ee{\end{equation}}
\def\bea{\begin{eqnarray}}
\def\eea{\end{eqnarray}}

\title{Arguments against a dominantly hadronic origin of the VHE radiation from
the supernova remnant RX J1713-3946}

\author{ R. Plaga \\ 
Franzstr. 40, 53111 Bonn, Germany\footnote{E-mail address: rainer.plaga@gmx.de}}
\begin{document}
\maketitle
\begin{abstract}
The flux
of photons above 1 TeV from
the direction of the centre and a 
cloud in the western part of the nearby southern supernova remnant (SNR) RX J1713.7-3946
is calculated in the ``hadronic scenario'' that aims
to explain the intense VHE radiation from this remnant
with the decay of $\pi_0$ pions produced in nuclear collisions.
The expected flux from its centre is found to
fall short by about factor
40 from the one observed by the HESS collaboration.
This discrepancy presents a serious obstacle to the ``hadronic scenario''.
The theoretically expected flux from
the molecular cloud 
exceeds the one observed by HESS by at least a factor 3.
While the size of this discrepancy might still seem acceptable in the face
of various theoretical uncertainties, the result strongly
suggests a strict spatial correlation of the cloud with an excess
of TeV $\gamma$ radiation.
The observational lack of such correlations in the remnant
reported by HESS is another counter argument
against the hadronic scenario. In combination these arguments
cannot be refuted by choosing certain parameters for the total
energy or acceleration efficiency of the SNR.
\end{abstract}


\section{Introduction}
\label{intro}
The nearby supernova remnant (SNR) RX J1713.7-3946 is a brighter
source of TeV photons than any SNR in the northern sky and, together with the SNR
RX J0852.0-4622, may well be the VHE brightest SNR
in the entire sky (Aharonian et al., 2005). 
Two alternative scenarios were proposed for the dominant origin of the TeV radiation from
this source by Aharonian et al.(2005)\footnote{
Aharonian et al. (2005) argue that ``mixed'' scenarios, in which
both electrons and protons contribute to a major part of the total emission seem
unlikely because they require fine tuning of some parameters.}. 
In the {\it electronic scenario} it is due to inverse Compton
scattering of energetic electrons mainly on photons of the microwave background
radiation.  In the {\it hadronic scenario} it is due to the decay of neutral
pions that were produced in collisions of energetic protons gas ambient to the remnant.
Conclusive evidence for the latter scenario would be of momentous importance
for the theory of cosmic-ray origin, because it would constitute the first
direct evidence
for the acceleration of protons in SNRs.
However, it is my purpose here to present two serious problems with a 
hadronic scenario for RX J1713.7-3946.

\section{The basic argument}
RX J1713.7-3946 is probably the remnant of a supernova in A.D. 393 at
a distance D $\approx$ 1 kpc (Cassam-Chenai et al., 2004). 
Presently it has a very roughly spherical shape with a radius 
r$_{\rm SNR}$ $\approx$ 0.5$^{o}$ (8.6 pc).
While it mostly expands into a void that was probably blown by the wind
of the supernova progenitor, it has recently hit a complex of molecular clouds in its
western part. In this direction the ambient density has been measured to be
be about 300/cm$^3$, while in the centre the ambient density has been observationally constrained
to $<$ 0.02/cm$^3$ (Cassam-Chenai et al., 2004)\footnote{Recently it has been claimed
that this upper limit can be avoided by assuming an extremely low electron temperature 
below 0.1 keV in the remnant, due to efficient particle acceleration (Takahashi et al., 2007). 
However, the measurements of Cassam-Chenai et al.(2004) find an electron temperature of 
about 0.6 keV in the central region of RX J1713-3946 (with low statistical significance).
Moreover, temperatures below 0.1 keV have not been found in any young SNR. 
The remnants with a claimed evidence for an
``acceleration cooling effect'' all  
have electron temperatures above 0.4 keV (Decourchelle et al., 2000).}.
In the hadronic scenario the TeV emissivity within
a cosmic-ray accelerator is directly
proportional to the amount of ambient ``target material''(Aharonian et al., 1994).
As most of the baryonic matter within RX J1713.7-3946 is clumped, the emissivity is
then expected to be strong in the direction of the molecular clouds
in the western part of the remnant and quite weak near the remnant's centre.
Yet, observations find a flux from its centre
that is only slightly weaker than 
the one from the western region.
Further there is no apparent spatial correlation of the flux  
with the molecular clouds (Aharonian et al., 2005).
These discrepancies between theoretical expectation and observation
constitute my basic arguments against a hadronic origin of the
VHE radiation from RX J1713.7-3946.
In the following they are refined and quantified, in the latter case for the example of one
well studied molecular cloud.

\section{Predictions for the VHE flux from $\pi_0$ decay}
\subsection{General parameterization}
I use a simple parameterization of Aharonian et al. (1994) to predict
the integral flux of $\gamma$-rays above an energy E:
\begin{equation}
\rm
F(\geq E) \approx 9 \cdot 10^{-12} \left( \theta \over 0.1 \right) \left(E \over 
{1 TeV}\right)^{-1.1}
\left(E_{SN} \over {10^{51} erg}\right) \left(D \over 1 kpc\right)^{-2} 
\left(n \over 1 cm^{-3}\right) cm^{-2} sec^{-1}.
\label{flux}
\end{equation}
Here $\theta$ is the fraction of the supernova explosion energy E$_{\rm SN}$ that
has been converted to cosmic rays with a power-law spectrum up to energies much
larger than E with an index of -2.1\footnote{This index is expected theoretically
and agrees with the index of the TeV spectrum of RX J1713.7-3946 within the errors up
to an exponential cutoff at about 20 TeV (Aharonian et al., 2005)}.
I choose the standard value for $\theta$, derived from a requirement that
SNRs are the main source of Galactic cosmic rays (Gaisser, 1990). I will
come back to the question if a different value might be chosen in the concluding 
section \ref{concl}.
D is the distance to the remnant and n is the ambient density of baryonic matter within
the source. This rough treatment is not expected to yield predictions with 
a precision of better than 10 $\%$. Therefore the inclusion of effects with
a small effect on the integral flux above 1 TeV, like e.g. the high-energy cutoff
of the observed spectrum (Aharonian et al., 2007) have been neglected.

\subsection{VHE radiation from the centre of the remnant}
The centre of the remnant contains no molecular clouds.
Because a strong shock must have passed and strongly heated this area,
the upper limit on the ambient gas density from the absence of thermal
X-ray radiation (Cassam-Chenai et al., 2004) is generally thought to be reliable
in this region.
On this basis, an upper limit 
on the integral flux above 1 TeV from the centre of RX J1713.7-3946
within the ``resolution radius'' of r$_{\rm S}$ $\approx$ 4.8$^{'}$ for the HESS 
measurements (that contains 68 $\%$ of all events from
a point source) is given by:
\begin{eqnarray}
\rm
F_{theory-centre}(\geq 1 TeV) \leq 7 \cdot 10^{-15} \left( \theta \over 0.1 \right)
\left(E_{SN} \over {10^{51} erg}\right) 
\nonumber
\\
\rm
\left(D \over 1 kpc\right)^{-2} 
\left(n \over 0.02 cm^{-3}\right)
\left( r_{SNR}/r_S \over 6.3\right)^{-2} cm^{-2} sec^{-1}
\label{flux_c}
\end{eqnarray}
\\
This equation is derived from eq.(\ref{flux}) by multiplying it
with the last factor that specifies the relative solid angle of the central
region relative to the total solid angle of the
remnant.
From fig.2 and table 2 of the recent publication of deep HESS
observations of RX J1713.7-3946 
(Aharonian et al., 2007) I find for the observed integral flux above
1 TeV from the direction of the centre of the remnant within a
resolution radius:
\begin{equation}
\rm
F_{observed-centre}(\geq 1 TeV) \approx 2.9 \cdot 10^{-13} cm^{-2} sec^{-1}
\label{flux_o} 
\end{equation}
The observed flux is larger by a factor of more than 40 than
the upper limit on the theoretically expected from $\pi_0$
decay. A hadronic origin of this radiation is improbable
on the basis of this discrepancy alone.

\subsection{VHE radiation from the direction of ``cloud C''}
Based on deep observations of RX J1713.7-3946 in the  
mm-wave and X-ray spectral region with various radio and X-ray telescopes 
it is generally thought that the shock wave of the remnant recently ran
into a complex of molecular clouds, identified by
their CO emission and optical absorption, 
in its western part (Cassam-Chenai et al., 2004; Hiraga et al., 2005).
The four major identified CO peaks (labelled ``A-D'' by Fukui at al.(2003)) are all located
on top of prominent X-ray features, suggesting that the
dense molecular gas is being impacted by blast waves and its
surface becomes bright in X-ray emission (Fukui et al., 2003).
Shock waves slow down fast and the cooling time scale
can become very short within dense clouds. Therefore the upper limits on the ambient
density from the absence of thermal X-ray radiation do not
apply for this region of the remnant (Cassam-Chenai et al., 2004).
\\
``Cloud C'' is an isolated molecular cloud, well within the south-western
rim of the remnant (Fukui et al., 2003). 
Its radius 
is about r$_{\rm c}$=3$^{'}$ (0.9) and I
idealize it as spherical.
Its coincidence with a 
region of enhanced X-ray brightness and a broad-line component
of CO emission are strong observational indications that this
cloud has interacted with the shock wave of RX J1713.7-3946 (Hiraga et al., 2005).
Both an absorption of X-ray spectra in the SW part of the remnant
and the intensity of the CO line allow to derive an additional absorption column density
at the position of the cloud 
of about N$_{\rm H}$=6 $\cdot$ 10$^{21}$ cm$^{-2}$ (from fig. 4 of
Cassam-Chenai et al. (2004) and fig.2 of Fukui et al. (2003), respectively).
This translates to an ambient density
within cloud C of about  n = N$_{\rm H}$/(2 r$_c$) $\approx$ 1100/cm$^3$.
The mass of cloud C is thus about 70 M$_{\odot}$.
The expected integral flux above 1 TeV from the direction of the
cloud within the resolution radius of 4.8 $^{'}$is then expected to be:
\begin{eqnarray}
\rm
F(\geq E)_{theory-cloud} = 1.4 \cdot 10^{-12} \left(\theta \over 0.1\right) 
\left(E \over {1 TeV}\right)^{-1.1}
\left(E_{SN} \over {10^{51} erg}\right) 
\nonumber
\\
\rm
\left(D \over 1 kpc\right)^{-2} 
\left(n \over 1100 cm^{-3}\right) 
\left( i \over 1/5 \right) \left( (r_{SNR}/r_c) \over 10 \right)^{-3} \rm cm^{-2} sec^{-1}
\label{flux_cloud}
\end{eqnarray}
This equation is a generalization of eq.(\ref{flux}) to estimate
the expected flux of $\gamma$ rays from a molecular cloud interacting
with the SNR. The last factor takes into account that
if the cosmic-ray density within the cloud is the same
as within the rest of the remnant (``full immersion'') 
the total fraction of cosmic rays interacting with the cloud
is the volume fraction of the cloud relative to the total remnant.
The factor ``i'' is an ``immersion factor'' that parameterizes the volume fraction of the
cloud that is immersed in the mean hadronic cosmic-ray density within the
remnant. 
It is difficult to determine i because the diffusion coefficient for
cosmic rays within the remnant is poorly known.
In the following I try to obtain a conservative 
{\it lower limit} on i.
\\
I assume that cosmic-rays 
are not accelerated 
within the dense cloud, due to a slower shock speed.
The entry into and propagation within the cloud is assumed
to be diffusive. This assumption seems justified by observational evidence that the
turbulent and magnetic energy density are practically equal
within molecular clouds (Crutcher, 1999).
The diffusion coefficient D$_ {\rm intercloud}$ of cosmic rays within the cloud
is estimated to have the standard interstellar value (Gaisser, 1990) scaled
with the ratio r of interstellar to intercloud magnetic field
strength. Assuming r $\approx$ 10 for cloud C (Crutcher, 1999)
I find D$_ {\rm intercloud}$ $\approx$ 5 $\times$ 10$^{29}$ cm$^2$/sec at an energy of 10 TeV.
I conservatively neglect diffusion into the cloud
from within the remnant and
assume that the diffusion of protons and nuclei into the 
cloud takes place exclusively from the region upstream of the shock, the {\it precursor}. 
Within the precursor I assume Bohm diffusion with a 
magnetic field of 3 $\mu$G i.e. a diffusion constant D$_p$ = c r$_{\rm L}$/3 where
r$_{\rm L}$ is the Larmor radius.
The measured spectrum from
RX 1713.7-3946 shows a highly significant exponential cutoff at an energy 
E$_{\rm max}$ $\approx$ 20 TeV (Aharonian et al., 2007).
My estimate for D$_p$  
yields a maximum proton (or electron) energy 
E$_{\rm max}$ $\approx$ 50 TeV
within the standard theory of shock-wave acceleration
(Gaisser, 1990)
\footnote{
An acceleration time of 1600 years and a shock speed v$_{\rm s}$ $\approx$ 5000 km/sec
were assumed.}.
This value is similar to 
the observed value of the exponential cutoff
of the spectrum, thus confirming the consistency
of my parameter choices.
I assume a precursor width of l$_{\rm CR}$ $\approx$ D$_p$/v$_{\rm s}$ 
(Malkov et al., 2005) and a time
period t$_{\rm p}$ during which the particles diffuse into the cloud 
of t$_{\rm p}$ = l$_{\rm CR}$/v$_{\rm s}$.
\\
The diffusion coefficient in the 
precursor D$_p$ $\approx$ 2 $\times$ 10$^{26}$ cm$^2$/sec is about a factor
5000 smaller than the one in the cloud D$_{\rm intercloud}$. We therefore assume that
the diffusion into the cloud is limited by the replenishment of
cosmic rays from the precursor which is determined by D$_p$.
A depth of extraction x$_{\rm p}$ of ambient cosmic rays with
an energy E from the
precursor into the cloud is calculated 
by the plane source solution Fick's law (Gaisser, 1990).
D$_p$ and t$_p$ are then expressed by the expressions explained 
in the present paragraph and one obtains:
\begin{equation}
\rm
x_p > \sqrt{D_p t_p} = {\rm 0.06 pc}  \left(E \over {\rm 10 TeV}\right) \left( {\rm v_s} \over 5000 {\rm km/sec}\right)^{-1}
\label{pene}
\end{equation}
A lower limit on ``immersion factor'' is then given as:
\begin{equation}
\rm
i > \left( r_c + x_p \over r_c \right)^3 - 1 \approx 1/5
\label{imm}
\end{equation}
for r$_c$ $\approx$ 0.9 pc and x$_p$ $\approx$ 0.06 pc.
\\
From fig.2 and table 2 of the recent publication of deep HESS
observations of RX J1713.7-3946 
(Aharonian et al., 2007) I find for the observed integral flux above
1 TeV from the direction of cloud C within a
radius of 4.8 $^{'}$ (within which 68 $\%$ of all events from the cloud are expected
(Aharonian et al., 2005))
\footnote{This approximates cloud C as an effective point source for the HESS observatory.}:
\begin{equation}
\rm
F_{\rm observed-cloud}(\geq 1 TeV) \approx 5 \cdot 10^{-13} {\rm cm^{-2} sec^{-1}}
\label{flux_o_c} 
\end{equation}
The theoretical {\it lower} limit (eq. (\ref{flux_cloud})) {\it exceeds} the observed
value (eq. (\ref{flux_o_c})) by about a factor 3.
This predicted flux would have made cloud C the brightest
TeV feature in RX J1713.7-3946 by far (about a factor 2 brighter than the
its brightest TeV feature in the NW of the remnant).
With other words: even under extremely conservative assumptions
about acceleration within and diffusion into molecular clouds overtaken by
a supernova blast waves, 
cloud C should be a very prominent and (for the HESS experiment)
effectively {\it point-like} feature
if the VHE radiation were of hadronic origin.
While a discrepancy of a factor 3 might still be acceptable in the
face of various theoretical uncertainties, the prediction of a 
point-like excess from cloud C with the hadronic model seems robust.
\\
Contrary to this expectation the HESS collaboration found ``no apparent
correlation between CO intensity and the HESS gamma-ray excess'' 
(quote from a collaboration member (Funk, 2007))
in general, and (in fig.17 in Aharonian et al.(2005)) no VHE excess at the position
of cloud C (located at an azimuth of $\approx$ 170$^o$) in particular.

\section{Conclusion}
\label{concl}
There are strict upper limits on thermal
X-ray radiation from the SNR RX J1713.7-3946 from 
observations of several satellites.
In the centre of the remnant these limits translate
into restrictive upper limits on gas density. I argued
that there is a factor 40 too little baryonic matter in its centre
to explain the observed TeV radiation 
with hadronic processes if
RX J1713-3946 is a typical accelerator of Galactic cosmic rays.
($\theta$ $\approx$ 0.1).
\\
In principle 
there is more than enough matter for this feat in the western part of the 
SNR. 
However, in order to avoid the strict upper limits on thermal
X-ray radiation this gas must be clumped in
dense molecular clouds, and exactly this is found from the 
spatial distribution of CO.
One would then expect that the TeV radiation is emitted mainly
from the direction of the clouds and again this is in contradiction
with observation.
\\
The two problems cannot both be solved by postulating a different acceleration
efficiency $\theta$ or supernova explosion energy E$_{\rm SN}$: increasing
either of these factors helps with the ``centre problem'' but aggravates the ``cloud problem''
and vice versa. The discrepancy further worsens if the cosmic-ray density should
be higher at the remnant's rim than at its centre, as might be expected
in the hadronic scenario.
These considerations practically rule out a mainly hadronic origin
of the VHE radiation from RX J1713.7-3946. This conclusion is
in disagreement with the one from Berezhko $\&$ V\"olk (2006) who argue
that the observed high energy $\gamma$-ray emission can be
mainly of hadronic origin. However, these authors do not take
advantage of the 
valuable morphological information that second generation Cerenkov arrays like
HESS and the X-ray satellites provide 
with high angular resolution. I base my arguments on exactly this information.
\\
The ``striking correlation between the X-ray and the gamma-ray image''
(quote from Aharonian et al. (2005))) is evidence in favour of
a leptonic origin of the radiation from RX J1713-3946. No model-independent
argument has been brought forward against it, yet.
The (relatively minor, i.e. smaller than a factor 2) disagreements between
the ``electronic scenario'' of Aharonian et al. (2005) and observations
could be due to idealizations they employ. 
The observational data in the X-ray and VHE spectral region
can be explained as purely of leptonic origin in the
following manner.
The intensity in the X-ray and VHE spectral region 
correlates quantitatively nearly perfectly 
throughout the SNR
(see e.g. fig. 16 in (Aharonian et al., 2005)).
The spectral properties are the same 
everywhere within the measurement error in the VHE region (see fig.14
in (Aharonian et al., 2005)) and X-ray region (see table 2 in
(Hiraga et. al, 2005)\footnote{Cassam-Chenai et al. (2004) find a
weak correlation between spectral hardness and intensity.}).
Porter et al. (2006) have proposed an energy distribution
for an electron population, a magnetic-field strength
and a detailed radiation-field energy distribution
that is shown to reproduce the observed total spectrum
from the radio up to the highest energies reasonably well.
The simple assumptions that the magnetic field and 
radiation-field energy distribution are the same throughout
the remnant and that the density of the electron
population correlates with the X-ray intensity 
will then be in satisfactory agreement with all data in the 
X-ray and VHE region.
While this model is probably too simplistic,
it serves to demonstrate that the radiation from RX J1713-3946 might
well be of purely leptonic origin. 
\section*{Acknowledgements}
I thank Werner Hofmann, Alvaro de Rujula and an anonymous referee
for critical remarks on a previous version of this
paper, that helped to improve it significantly.
\section*{References}

Aharonian, F.A., Drury, L.O'C., Voelk, G., 1994, A\&A 285, 645.
\\ 
Aharonian, F., et al., 2005, A\&A 449, 223.
\\
Aharonian, F., et al., 2007, A\&A 464, 235.
\\
Berezhko, E.G., V\"olk, H.J., 2006, A\&A 451, 981.
\\
Cassam-Chenai, G., Decourchelle, A., Ballet, J. et al., 2004, A\&A 427,199.
\\
Crutcher, R.M., 1999, ApJ 520, 706.
\\
Decourchelle, A., Ellison, D., Ballet, J., 2000, ApJ 543, L57.
\\
Fukui, Y., Moriguchi, Y., Tamura, K. et al., 2003, PASJ 55, L61.   
\\
Funk, S., 2007, astro-ph/0701471; 
Advances in Space Research (Proceedings COSPAR 2006), in print.
\\
Gaisser, T.K., {\it Cosmic Rays and Particle Physics}, (Cambridge University
Press, New York, 1990). 
\\
Hiraga, J.S. et al., 2005, A\&A 431, 953.
\\
Malkov, M.A., Diamond P.H., Sagdeev, R.Z., 2005, ApJ 624, L37.
\\
Porter, T.A., Moskalenko I.V., Strong A.W., ApJ 648, L29.
\\
Takahashi, T. et al., 2007, preprint astro-ph/0708.2002.


\end{document}